\documentclass[preprint,12pt,3p]{elsarticle}

\usepackage{amssymb}
\usepackage{amsmath}

\begin{document}

\begin{frontmatter}

\title{Gauge functions and Galilean invariance of Lagrangians}

\author{Z. E. Musielak and T. B Watson}
\address{Department of Physics, University of Texas 
at Arlington, Arlington, TX 76019, USA \\}

\begin{abstract}
A novel method to make Lagrangians Galilean invariant is developed. The method, based on 
null Lagrangians and their gauge functions, is used to demonstrate the Galilean invariance of 
the Lagrangian for Newton's law of inertia. It is suggested that this new solution of an old 
physics problem may have implications and potential applications to all gauge-based theories 
of physics. 
\end{abstract}

\end{frontmatter}

\section{Introduction}
  
Invariant equations of motion can be derived from Lagrangians that are not themselves 
invariant.  The best-known example is Newton's law of inertia, whose equation of motion 
is invariant with respect to the transformations that form the Galilean group of the metric 
[1], despite the fact that its standard Lagrangian is not [2,3].  This is possible because the 
Galilean transformations induce the Galilean gauge [3], which is either omitted [2] or 
removed by redefining the standard Lagrangian [3]. 
 
A novel method to enforce Galilean invariance on the Lagrangian is developed.  Using this 
method, a null Lagrangian can be constructed and its gauge function allows removing the 
Galilean gauge induced by the Galilean transformations.  After the Galilean gauge is removed, 
the Galilean invariant Lagrangian is obtained. This Lagrangian may still contain gauges 
introduced by the null Lagrangian; however all remaining gauges are also Galilean invariant.  
The developed method is general and may be extended to other physical problems that 
involve Galilean or Poincar\'e invariance of dynamical equations of classical and quantum 
mechanics.  

The main characteristics of a null Lagrangian (NL) are that it causes the Euler-Lagrange (E-L) 
equation to vanish identically, and that it can be expressed as the total derivative of a scalar 
function [4-7], called a gauge function [3,5].  This means that the NL can be added to any 
Lagrangian without changing the derivation of the original equation.  In general, the gauge 
function can be any continuous and differentiable function [4-7]; however, in this Letter, 
we construct an explicit gauge function of lowest orders in the dynamic variables.   

The developed method applies to second-order ordinary differential equations (ODEs), such 
as Newton's equations of dynamics, harmonic oscillators with and without damping, and 
other physical systems with similar equations of motion; the method can be generalized to 
partial differential equations (PDEs).  In the specific application presented in this Letter, we 
resolve the long-standing problem of the Galilean non-invariance of the Lagrangian for 
Newton's law of inertia via the construction of the explicitly Galilean invariant standard 
and null Lagrangians for this equation.

The NLs and their gauge functions are important in studies of Noether's [8,9] and other 
[10,11] symmetries in physical systems with given Lagrangians, in Carath\'eodory's theory 
of fields of extremals, and in integral invariants [5,6].  Moreover, the NLs play a role in 
studies of elasticity, where they physically represent the energy density function of materials 
[12].  There is a large body of literature on NLs (also called trivial Lagrangians [6,7]), and 
their applications (e.g., [13-15]).  It is shown that the obtained NLs can be applied to 
fundamental (Galilean invariant) equations of physics, and suggested that this approach 
may be extended to other gauge-based theories of modern physics.

\section{Lagrangians and gauge functions}

According to Newton's law of inertia, the motion of a body is always rectilinear and uniform 
with respect to an inertial frame of reference.  Using the Galilean group of the metric, Galilean 
observers associated with different intertial frames that are moving with respect to each other 
with constant velocities, are introduced.  The observers agree on description of physics in these 
frames and they use a Cartesian coordinate system ($x$, $y$, $z$) with time $t$ being the same 
in all interial frames (see Sect. 3). 

Then, the law of inertia for one-dimentional (along $x$) motion of a body in one inetrial frame 
can be expressed as   

\begin{equation}
\hat D x (t) = {{d^2 x (t)} \over {dt^2}} = \ddot x (t) = 0\ ,
\label{S2eq1}
\end{equation}
where $x(t)$ is a dynamical variable of the body in one inertial frame.  

It is known since the work of Lagrange in the 18th Century that the Lagrangian for this 
equation is 

\begin{equation}
L_{s} [\dot x(t)] = {1  \over 2} C_o \left [ \dot x (t) \right ]^2\ ,
\label{S2eq2}
\end{equation}
where $C_o$ is an arbitrary constant. We refer to this Lagrangian as standard because 
of its origin and dependence on the square of the dependent variable time derivative, 
which is a kinetic energy-like term if $C_o$ represents the mass of a moving body.   

Since the standard Lagrangian (SL) depends on the square of $\dot x (t)$, we construct 
two test-Lagrangians, one that  combines the dependent variable with its derivative or 
combines the dependent variable (or its derivative) with the independent variable, and 
the other that depends exclusively on either the dependent or independent variable.
We write these Lagrangians as 
\begin{equation}
L_{a} [\dot x(t), x(t), t] = C_1 \dot x (t) x (t) + C_2 \dot x (t) t + C_3 
x (t) t\ ,
\label{S2eq3}
\end{equation}
and
\begin{equation}
L_{b} [\dot x(t), x(t), t] = C_4 \dot x (t) + C_5 x (t) + C_6\ ,
\label{S2eq4}
\end{equation}
where $C_1$, $C_2$, $C_3$, $C_4$, $C_5$ and $C_6$ are constants as yet undetermined. 
The constants must have different physical dimensions to match the dimensions of $L_{a} 
[\dot x(t), x(t)]$ and $L_{b} [\dot x(t), x(t)]$ as that of $L_{s} [\dot x(t),x(t)]$.  This will 
be addressed when the constants are evaluated.

Defining $\hat {EL}$ to be the E-L equation operator, then $\hat {EL} (L_{n}) = 0$ is 
required for $L_{n} [\dot x(t), x(t)]$ to be a null Lagrangian\footnote{Our notation here 
is such that the subscript "$n$" denotes a "null" Lagrangian and is not to be taken as an index}. 
Adding $L_{a} [\dot x(t), x(t), t]$ and $L_{b} [\dot x(t), x(t), t]$, the condition $\hat {EL} 
(L_{a} + L_{b}) = 0$ is only valid if, and only if, $C_3 = 0$ and $C_5 = C_2$.  Then the 
NL is given by 
\begin{equation}
L_{n} [\dot x(t), x(t), t] = \sum_{i = 1}^{3} L_{ni} [\dot x(t), 
x(t), t]\ ,
\label{S2eq5}
\end{equation}
where $i$ = 1, 2 and 3, and the partial NLs are given by $L_{n1} [\dot x(t), x(t)] = C_1 
\dot x (t) x (t)$, $L_{n2} [\dot x(t), x(t), t] = C_2 [ \dot x (t) t + x (t) ]$ and $L_{n3} [\dot x(t)] 
= C_4 \dot x (t) + C_6$.  This is the most general NL that can be constructed by taking the 
lowest orders of the dynamical variable.

Adding the NL, Eq. (\ref{S2eq5}), to the SL, Eq. (\ref{S2eq2}), we obtain $L [\dot x(t), x(t), t] 
= L_{s} [\dot x(t)] + L_{n} [\dot x(t), x(t), t]$, or expressed in terms of the gauge function 
\begin{equation}
L [\dot x(t), x(t), t] = L_{s} [\dot x(t)] + {{d \Phi_n (t)} \over {dt}}\ .
\label{S2eq6}
\end{equation}
Using either $L [\dot x(t), x(t), t]$ or $L_{s} [\dot x(t)]$ the same equation 
of motion (see Eq. \ref{S2eq1}) is obtained.

The gauge function $\Phi_n (t)$ given in terms of the partial gauge 
functions is 
\begin{equation}
\Phi_n (t) = \sum_{i = 1}^3 \Phi_{ni} (t)\ , 
\label{S2eq7}
\end{equation}
where the partial gauge functions $\Phi_{ni} (t)$ correspond to the  
partial null Lagrangians $L_{ni} [\dot x(t), x(t)]$, and are defined as 
$\Phi_{n1} (t) = C_1 x^2 (t) / 2$, $\Phi_{n2} (t) = C_2 x (t) t$ and 
$\Phi_{n3} (t) = C_4 x (t) + C_6 t$.  It is not the aim of this Letter to 
demonstrate that the constructed $\Phi_n (t)$ is sufficient to make the 
SL given by Eq. (\ref{S2eq2}) Galilean invariant.
 
The existence of the standard Lagrangian for Eq. (\ref{S2eq1}) is guaranteed
by the Helmholtz conditions [16,17]; however, the existence of the NLs is
independent from these conditions because the NLs do not affect the derivation 
of the original equation.  In general, the problem of finding all NLs for a given 
equation has not yet been fully solved [6]; nevertheless, some progress has 
been made [5-15], and the presented results contribute to this progress.  

Typically, if the NLs are known they are omitted from the standard Lagrangian
[1] or removed by redefining this Lagrangian [2].  This is done in order to 
obtain a Galilean invariant Lagrangian free of the Galilean gauge [3].  Our 
novel result is that the NLs may be used to remove the unwanted Galilean 
gauges and make the standard Lagrangian Galilean invariant.

\section{Galilean invariance}  

In general, two coordinate systems may be rotated, translated and boosted relative to 
each other.  In Galilean space and time, all these transformations form the Galilean group 
of the metric.  This group is used to study Galilean invariance of classical [3,18] and 
quantum [19,20] physical systems.  The structure of this group is $G = \left[ T(1) \otimes 
R(3) \right] \otimes_s \left [ T(3) \otimes B(3) \right]$, where $T(1)$, $R(3)$, $T(3)$ and 
$B(3)$ are the subgroups of translation in time, rotations in space, translations in space and 
boosts, respectively.  The subgroups $T(1)$, $T(3)$ and $B(3)$ are Abelian Lie groups; 
however, the subgroup $B(3)$ is a non-Abelian Lie group. The direct product is denoted 
as $\otimes$, and $\otimes_s$ denotes the semi-direct product.  

Let $(x, t)$ be an inertial frame (see Sect. 2), and $(x^{\prime}, t^{\prime})$ be a another
intertial frame moving with respect to each other with velocity $v_0$ = const, and let the 
system's origins coincide at $t = t^{\prime} = t_0 = 0$.  Then the Galilean transformations 
relating these two systems are: $x^{\prime} = x - v_0 t$ and $t^{\prime} = t$.  In other 
words, a classical particle moving with the velocity $u = \dot x$ in the $(x, t)$ system has 
the velocity $u^{\prime} = \dot x^{\prime}$ in the $(x^{\prime}, t^{\prime})$.  These 
two velocities are related by the above Galilean transformation, so that $u^{\prime} = 
u - v_0$.

The solution of $\hat D x (t) = 0$ (see Eq. \ref{S2eq1}) can be written as $x (t) = a t + b$, 
where $a$ and $b$ are integration constants. Setting the following initial conditions $u 
(t)\vert_{t = 0} = \dot x (t)\vert_{t = 0} = u_0$ and $x (t)\vert_{t = 0} = x_0$, then $a 
= u_0$ and $b = x_0$, and the solution becomes $x (t) = u_0 t + x_0$.  It is shown 
below that using this solution some constants of the NL can be expressed in terms of 
$u_0$ and $x_0$.

\section{Galilean invariant Lagrangians}

Using Eq. (\ref{S2eq7}), we write the gauge function $\Phi_n (t)$ in the
explicit form 
\begin{equation}
\Phi_n (t) = {1 \over 2} C_1 x^2 (t) + C_2 x (t) t + C_4 x (t) + 
C_6 t\ .
\label{S4eq2}
\end{equation}
After a Galilean transformation ($x \rightarrow x^{\prime}$) with $t^{\prime} = t$, 
the transformed Lagrangian $L^{\prime} [\dot x^{\prime} (t), x^{\prime} (t), t]$ is 
given by
\begin{equation}
L^{\prime} [\dot x^{\prime} (t), x^{\prime} (t), t] = L_{s}^{\prime} 
[\dot x^{\prime} (t)] + {{d \Phi^{\prime}_{n} (t)} \over {dt}} + {{d} 
\over {dt}} \left [ \Phi^{\prime}_{Gs} (t) + \Phi^{\prime}_{Gn} (t) \right ]\ ,
\label{S4eq3}
\end{equation}
where
\begin{equation}
\Phi^{\prime}_n (t) = {1 \over 2} C_1 x^{\prime 2} (t) + C_2 x^{\prime} 
(t) t + C_4 x^{\prime} (t) + C_6 t\ ,
\label{S4eq4}
\end{equation}
is of the same form as $\Phi_n (t)$.  The Galilean invariant standard 
gauge function $\Phi^{\prime}_{Gs} (t)$ is 
\begin{equation}
\Phi^{\prime}_{Gs} (t) = C_0 \left [ x^{\prime} (t) + {1 \over 2} 
v_0 t \right ] v_0 \ .
\label{S4eq5}
\end{equation}
This gauge function is local and is the same as the phase 
of the wavefunction in the Schr\"odinger equation [19]; it 
corresponds to a projective unitary representation of the Galilean 
group of the metric [3].  The Galilean invariant null gauge function 
$\Phi^{\prime}_{Gn} (t) $ is also local and given by 
\begin{equation}
\Phi^{\prime}_{Gn} (t) = \left [ C_1 \left ( x^{\prime} (t) + {1 \over 2}
v_o t \right ) + C_2 t + C_4 \right ] v_o t\ .
\label{S4eq6}
\end{equation}

Both $\Phi^{\prime}_{Gs} (t)$ and $\Phi^{\prime}_{Gn} (t)$ are gauge 
functions because they give $L^{\prime}_{Gs} [\dot x^{\prime} 
(t), x^{\prime} (t), t]$ and $L^{\prime}_{Gn} [\dot x^{\prime} (t), x^{\prime} 
(t), t]$, respectively, which are the NLs. In order for $L [\dot x (t), x (t), t]$ and 
$L^{\prime} [\dot x^{\prime} (t), x^{\prime} (t), t]$ to be of the same form 
and Galilean invariant, $\Phi^{\prime}_{Gs} (t) + \Phi^{\prime}_{Gn} (t)$ 
must be either zero or constant; since the second case is more general, the 
condition $\Phi^{\prime}_{Gs} (t) + \Phi^{\prime}_{Gn} (t) = C$ is 
impose; the constants are evaluated using the initial conditions.

Using $x^{\prime} (t) = u^{\prime}_0 t + x^{\prime}_0$, where $u^{\prime}_0 
= u_0 - v_0$ and $x^{\prime}_0 = x_0$, we have $x^{\prime} (t) = ( u_0 - v_0 ) t 
+ x_0$, the following three constants can be evaluated $C = C_0 v_0 x_0$ 
\begin{equation} 
C_2 = - C_1 \left ( u_0 - {1 \over 2} v_0 \right )\ ,
\label{S4eq7}
\end{equation}
and
\begin{equation}
C_4 = - C_0 \left ( u_0 - {1 \over 2} v_0 \right ) - C_1 x_0\ .
\label{S4eq8}
\end{equation}
The constants remain the same in all inertial frames of reference. 

Both gauge functions $\Phi_n (t)$ and $\Phi^{\prime}_n (t)$ are Galilean 
invariant, which means that the null Lagrangians resulting from these functions 
are also Galilean invariant.  Thus, the Lagrangian in the $(x, t)$ frame 
is
\begin{equation}
L [\dot x (t), x (t), t] = L_{s} [\dot x (t)] + L_n [\dot x (t), x (t), t]\ ,
\label{S4eq9}
\end{equation}
where $L_{s} [\dot x (t)] = [ \dot x^2 (t)] / 2$, and $L_n [\dot x (t), x (t), t]$ 
is
\begin{equation}
L_n [\dot x (t), x (t), t] = C_1 \dot x (t) x (t) + C_2 \left [ \dot x (t) t 
+ x (t) \right ] + C_4 \dot x + C_6\ ,
\label{S4eq10}
\end{equation}
and the Lagrangian in the $(x^{\prime}, t^{\prime})$ frame becomes
\begin{equation}
L^{\prime} [\dot x^{\prime} (t), x^{\prime} (t), t] = L^{\prime}_{s} 
[\dot x^{\prime} (t)] + L^{\prime}_n [\dot x^{\prime} (t), x^{\prime} 
(t), t]\ ,
\label{S4eq11}
\end{equation}
where $L_{s} [\dot x^{\prime} (t)] = [(\dot x^{\prime})^2 (t)] / 2$ and 
$L^{\prime}_n [\dot x^{\prime} (t), x^{\prime} (t), t]$ can be written 
as
\begin{equation}
L^{\prime}_n [\dot x^{\prime} (t), x^{\prime} (t), t] = C_1 \dot x^{\prime} 
(t) x^{\prime} (t) + C_2 \left [ \dot x^{\prime} (t) t + x^{\prime} (t) \right ]
+ C_4 \dot x^{\prime} + C_6\ ,
\label{S4eq12}
\end{equation}
remains the same, which means that the Lagrangian is Galilean invariant.

\section{Physical implications}

In most previous studies of classical systems of physics, the null Lagrangians either 
never appeared or if they did were simply omitted [2,3] as unimportant in the 
derivation of equations of motion; the only known exception was a limited application 
of the null Lagrangians to elasticity [12].  Nevertheless, many mathematical aspects 
and applications of the null Lagrangians were investigated in the framework of the 
calculus of variations [4-7, 13-15].

The main result of this Letter is that the standard Lagrangian 
\begin{equation}
L_s [\dot x (t)] \rightarrow L^{\prime}_s [\dot x^{\prime} (t)] + L^{\prime}_{Gs} 
[x^{\prime} (t), t]\ , 
\label{S5eq1}
\end{equation}
which is not Galilean invariant, can only be made Galilean invariant if, and 
only if, the standard Lagrangian is supplemented by a null Lagrangian, which 
must also be Galilean invariant.  This can be written as  
\begin{equation}
L_s [\dot x (t)] + L_n [\dot x (t), x (t), t]  \rightarrow L^{\prime}_s [\dot x^{\prime} (t)] 
+ L^{\prime}_n [\dot x^{\prime} (t), x^{\prime} (t), t]\ ,
\label{S5eq2}
\end{equation}
where $L_s [\dot x (t)]$, $L_n [\dot x (t), x (t), t]$, $L^{\prime}_s [\dot x^{\prime} 
(t)]$ and $L^{\prime}_n [\dot x^{\prime} (t), x^{\prime} (t), t]$ are Galilean invariant
Lagrangians.

This shows the role that the null Lagrangians and their gauge functions play in making the 
standard Lagrangian for Newton's law of inertia Galilean invariant.  The obtained results 
demonstrate that there is only one general null Lagrangian, constructed to the lowest orders 
of the dynamical variable, that simultaneously remains Galilean invariant and also induces 
invariance in the standard Lagrangian.  The presented method to construct the Galilean 
invariant Lagrangian may be extended to different physical problems that require Galilean 
invariance of dynamical equations of classical or quantum mechanics as well as to physical 
problems that involve Poincar\'e invariance.  The obtained results also show that the null 
Lagrangians and their gauge functions are present in classical mechanics if the invariance 
of Lagrangians is required.

The presented results are relevant to studies of Noether [9,21,22], non-Noether [11,23,24]
and other [8,10,25] symmetries of Lagrangians and the equations of motion.  As demonstrated in
[9], the Noether symmetries of standard Lagrangians remain the same whether null Lagrangians
are added to them or not.  Similarly, standard and standard+null Lagrangians yield their two 
corresponding non-Noether symmetries in a unique way [23].  In general, Lagrangians posses 
less symmetry than the equations of motion resulting from them due to assumptions on which 
the Noether theorem [10,25] is based.  However, we postulate (without a proof) that the Galilean 
standard Lagrangian, not only allows deriving the Galilean invariant equation of motion [21], but 
also its underlying symmetries are identical with the symmetries of the resulting equation of 
motion.      

Symmetries of differential equations and their Lagrangians are related to Lie groups [4].  
For example, let $L (\dot q, q, t) \equiv L [\dot q (t), q (t), t]$ be a Lagrangian, $q$ be 
a generalized variable, $M$ be a configuration manifold, and $TM$ be a tangent bundle 
associated with this manifold, then $L: TM \rightarrow \mathcal{R}$, and $L$ is defined 
on $TM$ [26].  Let now $G$ be a manifold associated with a given Lie group $G$, so that 
$L: TG \rightarrow \mathcal{R}$.  Then, the Lagrangian $L (\dot q, q, t)$ remains also 
$G$-invariant but 'new' variational principles and 'new' Euler-Lagrange equations may be 
required [27,28].  The invariance of $L (\dot q, q, t)$ is important because it is strongly 
related to its Lie group [4].  The problem may also be reversed; for every known Lagrangian 
its invariance with respect to rotations, translations and boosts may indicate the presence 
of the underlying Lie group [29,30], which means that the group may be identified by 
investigating the Lagrangian invariance.  Moreover, the invariance of $L (\dot q, q, t)$ 
guarantees that the original equation derived from this Lagrangian preserves the same 
invariance [21].  This is the case for the Galilean invariant Newton's first equation of 
dynamics and its Galilean invariant standard and null Lagrangians derived in this Letter.  
In addition, Galilean invariance guarantees that the same Lie group that underlies the 
Newton law of inertia is also the group for the Lagrangians.

\bigskip\noindent
{\bf Acknowledgments}
We are indebted to three anonymous referees for their valuable suggestions that allow 
us to improve significantly the first version of this manuscript.  The authors also would 
like to thank L. C. Vestal and N. Davachi for checking our derivations and their comments 
on the manuscript.  This work was supported by the Alexander von Humboldt Foundation 
(Z.E.M.).
%


\end{document}